\documentclass[pre,twocolumn,twoside,superscriptaddress,showpacs,floatfix]{revtex4-1}
\usepackage{amsmath,graphicx,multirow,natbib,hyperref}
\hypersetup{colorlinks,citecolor=black,filecolor=black,linkcolor=black,urlcolor=black}
\usepackage[switch]{lineno}
\usepackage{amsmath}
\usepackage{amssymb}
\usepackage{empheq}
\usepackage[parfill]{parskip}
\usepackage[active]{srcltx}
\usepackage{color}
\usepackage{array}
\usepackage{booktabs}
\usepackage{amsfonts}
\usepackage{dsfont}
\usepackage{graphicx}

\begin{document}

\setlength{\parindent}{0.5cm}

\title{Swarmalators with thermal noise}

\author{Hyunsuk Hong}
\affiliation{Department of Physics and Research Institute of Physics and Chemistry, Jeonbuk National University, Jeonju 54896, Korea}
\affiliation{School of Physics, Korea Institute for Advanced Study, Seoul 02455, Korea}

\author{Kevin P. O'Keeffe}
\affiliation{Senseable City Lab, Massachusetts Institute of Technology, Cambridge, MA 02139, USA}

\author{Jae Sung Lee}
\affiliation{School of Physics, Korea Institute for Advanced Study, Seoul 02455, Korea}

\author{Hyunggyu Park}
\affiliation{Quantum Universe Center, Korea Institute for Advanced Study, Seoul 02455, Korea}

\begin{abstract}
We investigate a population of swarmalators, a mobile version of phase
oscillators that both sync in time and swarm through space.
We focus on a XY-type model of identical swarmalators running on a one-dimensional
ring and subject to thermal noise. We uncover four distinct collective states, some of
which capture the behavior of real-world swarmalators such as vinegar eels
and sperm. Among these, the most intriguing is the `mixed state', which
blends two of the other states. We present a comprehensive phase diagram from the Fourier mode analysis
with a high accuracy, which is in excellent agreement with numerical simulation results.
Our model serves as a tractable toy model for thermal systems that both self-synchronize and self-assemble
interdependently.
\end{abstract}

\maketitle

\section{Introduction}
Swarmalators, short for ``swarming'' oscillators,  are mobile phase oscillators that can both synchronize in time and swarm through space~\cite{ref:swarmalators}.
They were recently introduced to model the numerous systems in which synchronization and swarming co-exist and interact.
Examples are sperm~\cite{ref:sperms},
vinegar eels~\cite{ref:vinegareels}, magnetotactic bacteria,
starfish embroys~\cite{ref:livingcrystals},
Japanese Tree frogs~\cite{ref:Japanesefrogs},
Janus particles~\cite{ref:Janusparticles}, Quincke rollers~\cite{ref:Quinckerollers},
and robotic swarms~\cite{ref:robots}.

Research on swarmalators began to gain traction about  15 years ago when Iwasa and Tanaka introduced a universal model of chemotactic oscillators which produced diverse states~\cite{ref:chemo,ref:diverse}. O'Keeffe, Hong, and Strogatz later introduced a generalized Kuramoto model which produced five collective states that are commonly observed in nature~\cite{ref:swarmalators}. This swarmalator model is being used as a spring-board to further study the collective behavior of swarmalators. The effects of pinning \cite{ref:pinning}, forcing \cite{ref:forcing}, and various types of coupling (delayed \cite{ref:delay}, finite-range \cite{ref:finiterange} stochastic \cite{ref:stochastic},
mixed sign \cite{ref:coupling, ref:gourab, ref:adv, ref:pre-coupling})
as well as other phenomena/scenarios \cite{ref:ring,ref:mfl,ref:topological,ref:jimenz,ref:ha,ref:steven} have been explored.  For reviews of swarmalators, see \cite{ref:review1,ref:review2}.

Theoretical results on swarmalators are scarce due to the complexity of the system, which is characterized by nonlinear couplings of numerous degrees of freedom. For a system of $N$ particles, each with position $x \in \mathbb{R}^d$ and phase $\theta \in S^1$, the analysis of the  nonlinear coupled ordinary differential equations (ODEs) is challenging. What is missing in the swarmalator field is the ``right'' or natural toy model such as the Ising model for equilibrium phase transition study~\cite{ref:Ising} or the Kuramoto model for synchronization studies~\cite{ref:Kuramoto}, which captures the essential ingredient while remaining simple enough to solve.

This paper focuses on a simple one-dimensional ($1$D) swarmalator model~\cite{ref:1D,ref:1Dwithfrequency} which could fill this `modeling gap' in studies of swarmalation, as Verberk \cite{ref:verberck} has dubbed the interplay between \underline{swarm}ing and synchronizing oscil\underline{lations}. The model confines the swarmalators to run on a 1D ring, which makes it analytically tractable yet it still produces real-world behavior; its collective states capture the behavior of real-world 1D swarmalators, such as bordertaxic vinegar eels and sperm, and also capture the rotational analogue of 2D \& 3D swarmalators, such as magnetic Janus particles and dielectric Quinke rollers.

In a recent study~\cite{ref:1Dwithfrequency}, the 1D swarmalator model with ``quenched'' disorder
was explored, wherein the natural frequencies of each swarmalator were
chosen randomly  and then fixed for all times (nonidentical oscillators).
The authors utilized a ``toroidal''
Ott-Antonsen(OA) ansatz on a special submanifold of state space to identify four distinct long-lived states~\cite{ref:1Dwithfrequency}.
Meanwhile, systems in nature are typically influenced by their surrounding environment, corresponding to a heat bath,
which implies that natural systems are {\em stochastic} and affected by thermal noise.
In this context, it is pertinent to investigate the effects of  ``temporal/thermal'' disorder on
the collective properties of the system.

To this end, we introduce thermal noise into the 1D swarmalator model and explore its collective behavior.
Here, we restrict our analysis to the case of identical swarmalator only, leaving the generalization to nonidentical
ones with thermal noise for future study, as this would significantly complicate the analysis.
We note that our case with identical oscillators can be also interpreted as a coupled XY model for a magnetic spin system~\cite{ref:XY}, which provides various interesting nonequilibrium steady states.

We analyse the model both numerically and analytically. It is well known that the analytic OA ansatz fails in the presence of
thermal noise~\cite{ref:OA,ref:KS,ref:HJHP,ref:Pikovsky}. 
Consequently, the Fourier mode analysis does not close with a finite number of terms, which leads 
to an infinite hierarchy of coupled equations. Nevertheless, as the contributions from higher modes become increasingly negligible to the synchronization order parameters, the coupled equations can be suitably truncated and solved numerically to evaluate the order parameters highly accurately. Our results compare very well with numerical
simulations by integrating the equations of motion directly. We also analytically derive the instability condition for the completely disordered phase and employ a perturbation theory in order to reach out to a partially ordered phase. Our analysis is consistent with
numerical results.


\section{Swarmalators on a 1D ring}

We consider a population of $N$ coupled identical swarmalators on a 1D ring, of which the dynamics are governed by
\begin{align}
    \dot{x_i} &= \frac{J}{N} \sum_{j=1}^N \sin(x_j - x_i) \cos(\theta_j - \theta_i) + \xi_i^{x}, \label{eq:x} \\
    \dot{\theta_i} &= \frac{K}{N} \sum_{j=1}^N \sin(\theta_j - \theta_i ) \cos(x_j - x_i ) + \xi_i^{\theta}, \label{eq:theta}
\end{align}
where $x_i$ and $\theta_i$ denote the position and phase of the $i$th swarmalator, respectively, and are both
periodic with a period of $2\pi$. One may add an identical natural frequency $\omega$ for phase in Eq.~\eqref{eq:theta},
but it can be eliminated by taking a simple transformation of $\theta_i\rightarrow \theta_i + \omega t$. Likewise, the identical directed velocity $v$ can be also set to zero without loss of generality.
$\xi_i^{x}$ and $\xi_i^{\theta}$ represent thermal noise with zero mean, and characterized by
\begin{align}
\langle \xi_i^{x}(t) \xi_j^{x}(t') \rangle &= 2D_{x}\delta_{ij}\delta(t-t'), \label{eq:xi_x} \\
\langle \xi_i^{\theta}(t) \xi_j^{\theta}(t') \rangle &= 2D_{\theta}\delta_{ij}\delta(t-t'),
\label{eq:xi_theta}
\end{align}
where $D_x$ and $D_{\theta}$ denote the ``temperature'' of the heat bath for $x$ and $\theta$ variables, respectively.
The two noises are independent of each other, i.e.~$\langle\xi_i^x\xi_j^\theta\rangle =0$. $J$ and $K$ are the coupling parameters.

This 1D swarmalator model has been previously studied in the absence of thermal noises for identical swarmalators~\cite{ref:1D} and  for random nonidentical ones~\cite{ref:1Dwithfrequency}.
The novelty of this paper lies in the presence of thermal noises without quenched disorder for nonidentical swamalators.

Equation~\eqref{eq:theta} models position dependent synchronization, wherein the sine terms induce the swarmalators to reduce their phase difference for $K>0$, while the cosine term implies that the synchronization tendency is stronger for oscillators which are closer in space.  Equation~\eqref{eq:x} is the mirror-image of Eq.~\eqref{eq:theta}, which models phase-dependent swarming or aggregation
for $J>0$.
This can be thought as synchronization occurring on a torus, as both position and phase are circular variables $(x_i, \theta_i) \in \mathbb{T}^{1}$. In this sense, it can be referred to as a generalized version of the thermal Kuramoto model~\cite{ref:KS} for two coupled populations of identical oscillators.

This model can be also seen as a nonequilibrium generalization of the widely-recognized XY model for a magnetic spin system~\cite{ref:XY}, wherein
spins possess two degree of freedoms, $x$ and $\theta$, that are coupled.
For the special case of $J=K$ and $D_x =D_\theta$, the dynamics described by Eqs.~\eqref{eq:x} and \eqref{eq:theta} is equivalent  to that of an equilibrium (EQ) system in contact with a thermal reservoir of temperature $D_x $ with the Hamiltonian
\begin{align}
{{H}}=-\frac{J}{2N} \sum_{{i,j=1}}^N
\cos(x_j-x_i)\cos(\theta_j - \theta_i)~,
\label{eq:Hamiltonian}
\end{align}
where an EQ (magnetic) phase transition is expected as $J$ varies.
For $J\neq K$, the Hamiltonian structure is lost and the system can not relax to an equilibrium state. This is a typical nonequilibrium (NEQ) situation in the presence of external non-conservative forces, where various NEQ (magnetic) phases can emerge.
For $D_x\neq D_\theta$, this model may describe a heat engine under the thermal gradient.

Yet another way to interpret the model is as a 1D Vicsek-type model~\cite{ref:Vicsek} where $\theta$ corresponds to an orientation, as opposed to an internal phase variable.
\begin{figure}
   \centering
    \includegraphics[width=0.95\columnwidth]{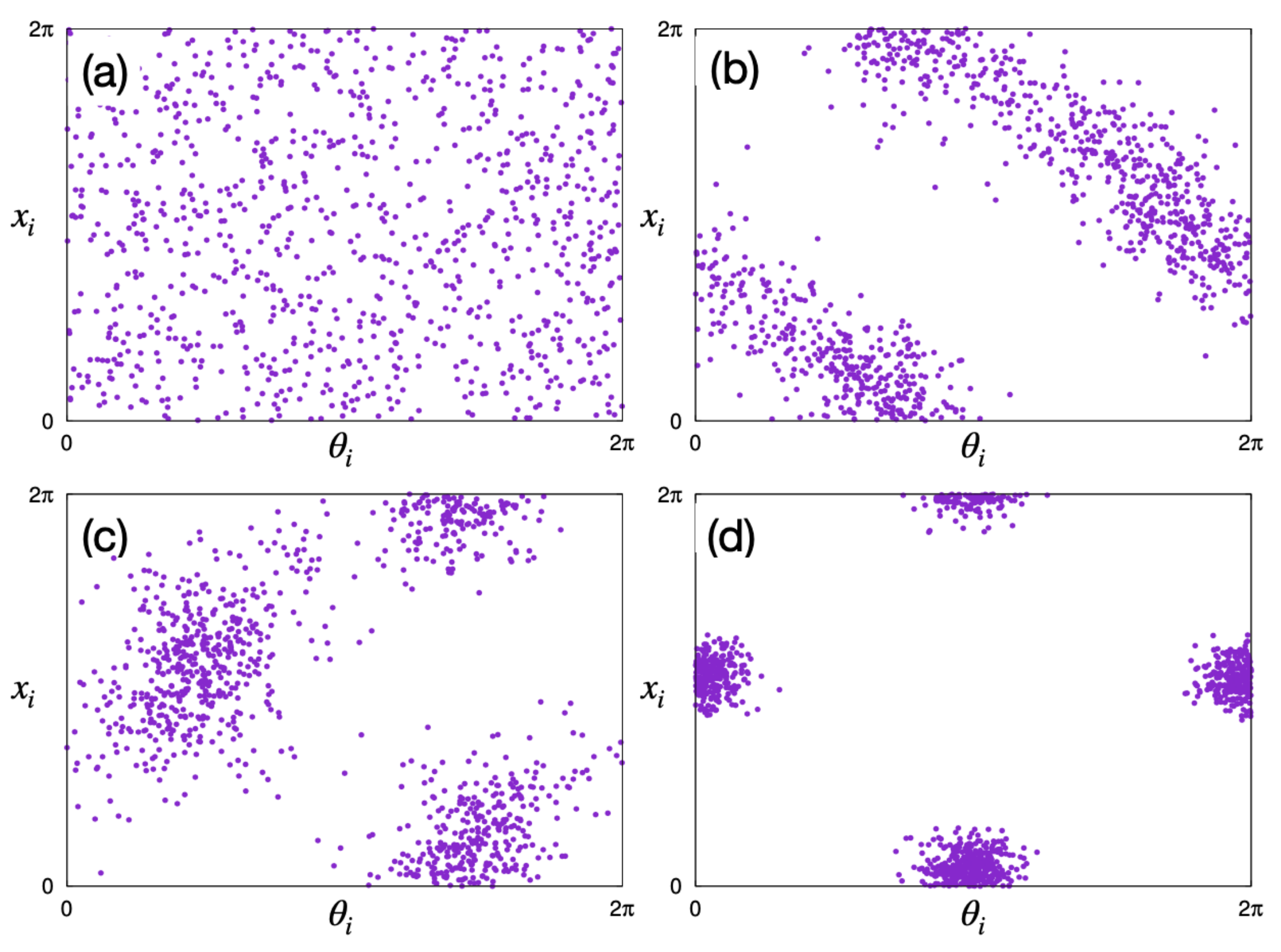}
    \caption{(Color online) Scatter plots of the four states: (a) $(0,0)$ state
for $(J, K)=(0,0)$; (b) $(S, 0)$ state for $(J, K)=(2.09, 0.31)$;
(c) $(S_1, S_2)$ state for $(J, K)=(2.04, 0.36)$; (d) $(S,S)$ state for
$(J, K)=(2.52, 2.07)$. We set the temperature $D_{x}=D_{\theta}=0.15$ and the number of swarmalators $N=1000$.
    }
    \label{fig:scatterplot_states}
\end{figure}

\section{Order parameters}

In order to gain insight into the order parameters characterizing collective behaviors,
numerical simulations are carried out by integrating Eqs.~\eqref{eq:x} and \eqref{eq:theta} for various parameter values.
Four distinct steady states were identified, and their typical states are visualized
in  Fig.~\ref{fig:scatterplot_states}; (a) uniformly distributed, (b) linearly correlated between $x$ and $\theta$,
(c){mixed} between (b) and (d), and (d) two well-defined isotropic clusters separated by a distance of $\pi$ in both directions.

One may consider the Kuramoto synchronization (or XY magnetic) order parameters~\cite{ref:Kuramoto} for each variable $x$ and $\theta$ as
\begin{align}
Z_{x} &\equiv R_{x}e^{i\Theta_x} = \frac{1}{N} \sum_{j=1}^{N} \langle e^{i x_j}\rangle, \label{eq:Rx} \\
Z_{\theta} &\equiv R_{\theta}e^{i\Theta_{\theta}} = \frac{1}{N} \sum_{j=1}^{N} \langle e^{i \theta_j}\rangle, \label{eq:Rtheta}
\end{align}
where $R_{x}$ and $R_{\theta}$ measure the magnitude of synchronization  (real and non-negative) in $x$ and $\theta$,
respectively, and $\Theta_{x}$ and $\Theta_{\theta}$ are their mean angles (real). $\langle\cdots\rangle$ denotes
the thermal (noise) average.
However, they are not effective order parameters, as they vanish in all four phases.
This is due to the special coupling structure in Eqs.~\eqref{eq:x} and \eqref{eq:theta},
which are invariant under the ``$\pi$'' transform of $x_j\rightarrow x_j+\pi$ and $\theta_j\rightarrow \theta_j+\pi$ for
any $j$, as noted  in Refs.~\cite{ref:1D} and \cite{ref:1Dwithfrequency}.
It implies that the dynamics of any $(x,\theta)$ state is equivalent to that of the $(x+\pi, \theta+\pi)$ state.
In particular, the number of swamalators in each cluster in (c) and (d) appears to be equal, as is expected for large $N$ in
accordance with combinatorial theory. Thus, the contributions from each cluster cancel out, leading to
$Z_x=Z_\theta=0$  in the $N\rightarrow\infty$ limit. Likewise, the same conclusion can be drawn for (b).


Thus, it is necessary to explore alternative order parameters that can be used to quantify
the correlation between two variables, which were proposed previously in~\cite{ref:1D} and \cite{ref:1Dwithfrequency},
as
\begin{align}
W_{+} &\equiv S_{+} e^{i\Phi_{+}} = \frac{1}{N}\sum_{j =1}^N \langle e^{i (x_j+\theta_j)}\rangle, \label{eq:Sp} \\
W_{-} &\equiv S_{-} e^{i\Phi_{-}} = \frac{1}{N}\sum_{j =1}^N \langle e^{i (x_j-\theta_j)}\rangle,\label{eq:Sm}
\end{align}
where $S_{\pm}$ is real and non-negative by definition and $\Phi_{\pm}$ is real.
This particular form of the correlation measure can be inferred from the linearity of correlations between $x$ and $\theta$
with slopes of $\pm 1$, as seen in Fig.~\ref{fig:scatterplot_states}.  With these order parameters, the four distinct steady states can be clearly distinguished as follows:
(a) Disordered (incoherent) state with $(S_+,S_-)=(0,0)$; (b) {\em Phase wave} state with $(S,0)$ or $(0,S)$, where the value of $S$ depends on
coupling parameters. The $(S,0)$ and $(0,S)$ states are equally probable, analogous to the magnetically ordered up and down states in
the Ising model; (c) {\em Mixed} state with $(S_1, S_2)$ or $(S_2,S_1)$, where $S_1\neq S_2$ and both are finite.
Finally, (d) Synchronized (ordered) state with $(S,S)$, where both correlations in Eqs.~\eqref{eq:Sp} and \eqref{eq:Sm}
are equal in magnitude.

\section{Coordinate transformation}

It is convenient to make coordinate transformations as
\begin{align}
X_i\equiv x_i+\theta_i~, ~\textrm{and}~~Y_i \equiv x_i-\theta_i~,
\end{align}
which imply that the new variables $X$ and $Y$ are both periodic with a period of $4\pi$.
Subsequently, the dynamic equations in Eqs.~\eqref{eq:x} and \eqref{eq:theta} can be simplified to
\begin{align}
\dot{X}_i &= J_{+} S_{+}\sin(\Phi_{+}-X_i) + J_{-} S_{-}\sin(\Phi_{-}-Y_i) + \xi_i^X, \label{eq:X} \\
\dot{Y}_i &= J_{-} S_{+}\sin(\Phi_{+}-X_i) + J_{+} S_{-}\sin(\Phi_{-}-Y_i) + \xi_i^Y, \label{eq:Y}
\end{align}
with $J_{\pm}=(J\pm K)/2$, and $S_{\pm}$ and $\Phi_{\pm}$ are defined in Eqs.~\eqref{eq:Sp} and \eqref{eq:Sm}.

The new noises, $\xi^{X}_i(t)$ and $\xi^{Y}_i(t)$, are defined as
\begin{align}
\xi^{X}_i = \xi^{x}_i+\xi^{\theta}_i~, ~\textrm{and}~~\xi^{Y}_i = \xi^{x}_i-\xi^{\theta}_i, \label{eq:xi_Y}
\end{align}
that are {\em not} independent of each other in general.
Simple calculations lead to $\langle \xi_i^{X}\rangle=\langle \xi_i^{Y}\rangle=0$ and
\begin{align}
\langle \xi_i^{X}(t) \xi_j^{X}(t') \rangle &= 2D_{X}\delta_{ij}\delta(t-t'), \label{eq:xi_X} \\
\langle \xi_i^{Y}(t) \xi_j^{Y}(t') \rangle &= 2D_{Y}\delta_{ij}\delta(t-t'), \label{eq:xi_Y} \\
\langle \xi_i^{X}(t) \xi_j^{Y}(t') \rangle &= 2D_{XY}\delta_{ij}\delta(t-t'), \label{eq:xi_XY}
\end{align}
where
\begin{align}
&D_{X}=D_{Y}=D_{x}+D_{\theta}\equiv D~, \label{eq:D}\\
&D_{XY}=D_x-D_\theta\equiv d~.\label{eq:d}
\end{align}
The correlation between the two noises $D_{XY}$ disappears when the two temperatures are equal ($D_x=D_\theta$),

It is worth noting that the dynamic equations in Eqs.~\eqref{eq:X} and \eqref{eq:Y} are invariant
under the transformation of $X_i\rightarrow X_i+2\pi$ or $Y_i\rightarrow Y_i+2\pi$ for any $i$, implying
that the variables $X$ and $Y$ can be treated as periodic with a period of  $2\pi$,
instead of $4\pi$, if the initial distributions possess the same periodicity.

When $J=K$ ($J_-=0$) with $D_x=D_\theta$ ($d=0$),
the dynamic equations decouple completely, leading
to the EQ XY model with two decoupled degree of freedoms for $X$ and $Y$,
which can be solved exactly.
By varying $J$, we find that the EQ magnetic transition occurs at $J_c=2D_X=4D_x$ from the disordered state $(0,0)$ (a),
directly to the ferromagnetic ordered state $(S,S)$ (d) without an intermediate phase wave or mixed state.
Intermediate states (b) and (c) may emerge only when the system is driven out of EQ, either by an external  non-conservative force ($J\neq K$) or a thermal gradient caused by multiple reservoirs ($D_x\neq D_\theta$).

\section{Fourier mode analysis}

In the $N\rightarrow\infty$ limit,
the system state can be described by a  continuous probability density function (PDF)
$\rho (X,Y,t)$ at time $t$ in terms of new variables $X$ and $Y$. As discussed in the preceding section,
these variables can be treated as periodic with a period of $2\pi$, thus their range is both set to $(0, 2\pi)$
with the normalization condition as $\int_0^{2\pi}\int_0^{2\pi} dX dY \rho(X,Y, t)=1$.

The Fokker-Planck equation~\cite{ref:Risken}, derived from
the dynamic equations of \eqref{eq:X} and \eqref{eq:Y}
with  Eqs.~\eqref{eq:D} and \eqref{eq:d}, is given by
\begin{align}
    \frac{\partial \rho}{\partial t}
= -\frac{\partial}{\partial X}{\cal J}_X -\frac{\partial}{\partial Y}{\cal J}_{Y} \label{eq:FP}
\end{align}
where the probability currents ${\cal J}_X$ and ${\cal J}_Y$ are
\begin{align}
    {\cal J}_X &=(J_+ F_{+} + J_{-} F_{-})\rho -D\frac{\partial \rho}{\partial X}
    -d\frac{\partial \rho}{\partial Y}\\
    {\cal J}_Y &=(J_- F_{+} + J_{+} F_{-})\rho -D\frac{\partial \rho}{\partial Y}
    -d\frac{\partial \rho}{\partial X}~,
\label{eq:FP_Dxy}
\end{align}
with
\begin{align}
F_{+}\equiv S_{+}\sin(\Phi_{+}-X),~~F_{-}\equiv S_{-}\sin(\Phi_{-}-Y)~.
\end{align}

The Fourier series of the PDF  $\rho(X,Y,t)$ in $X$ and $Y$ with period $2\pi$ is given by
\begin{equation}
    \rho(X,Y,t)=\frac{1}{(2\pi)^2}\sum_{n,m=-\infty}^{\infty}
    {\alpha_{n,m}(t)}e^{in(X-\Phi_{+})}e^{im(Y-\Phi_{-})},
    \label{eq:rho_XY}
\end{equation}
where the extra factors $\Phi_+$ and $\Phi_-$ are inserted for convenience.
As the PDF should be real, $\alpha^{*}_{n,m}=\alpha_{-n,-m}$ holds for all $(n,m)$ and
the normalization condition yields $\alpha_{0,0}=1$.



The order parameters in Eqs.~\eqref{eq:Sp} and \eqref{eq:Sm} can be expressed in terms of Fourier coefficients as
\begin{align}
S_{+} &= \int dXdY~e^{i(X-\Phi_{+})}\rho(X,Y,t)  =\alpha_{-1,0}=\alpha^*_{1,0}, \label{eq:Sp_alpha10} \\
S_{-} &= \int dXdY~e^{i(Y-\Phi_{-})} \rho(X,Y,t) =\alpha_{0,-1}=\alpha^*_{0,1} \label{eq:Sm_alpha01}~.
\end{align}
It should be noted that, as $S_{\pm}$ is real, $\alpha_{1,0}$ and $\alpha_{0,1}$ must also be real.
Substitution of Eq.~\eqref{eq:rho_XY} into the Fokker-Planck equation of Eq.~\eqref{eq:FP}
yields time-dependent mode-coupled equations as
\begin{align}
&{\dot{\alpha}}_{n,m} -i(n\dot{\Phi}_++m \dot{\Phi}_-)\alpha_{n,m} \nonumber\\
&\quad =-[D(n^2+m^2)+2dnm]\alpha_{n,m}\nonumber\\
&\quad\quad+ \frac{1}{2}(nJ_{+}+mJ_{-})S_+(\alpha_{n-1,m} -\alpha_{n+1,m})
\nonumber\\
&\quad\quad+\frac{1}{2}(nJ_{-}+mJ_{+}) S_-(\alpha_{n,m-1}-\alpha_{n,m+1}) .
\label{eq:alphanm}
\end{align}
where $S_{+}=\alpha_{1,0}$ and $S_{-}=\alpha_{0,1}$.
These equations form an infinite hierarchy of coupled equations, which can not be reduced to
a finite number of equations unlike the Kuramoto model without thermal noise, where the OA ansatz holds.

\section{steady state solutions}\label{sec:sss}

The steady-state solutions (fixed points) $\rho^s (X,Y)$ can be obtained by setting $\dot\alpha_{n,m}=\dot\Phi_{\pm}=0$
in Eq.~\eqref{eq:alphanm}. There exists the trivial solution $\alpha_{n,m}=0$ for any pair of $(n,m)$ except
for $\alpha_{0,0}(=1)$, resulting in $S_+=S_-=0$ and $\rho^s_{\textrm{(a)}}=1/(2\pi)^2$ (disordered state (a)).

The solutions of the phase wave state  (b) can be obtained exactly by the choice of $S_+=0$ or $S_-=0$.
With $S_-=0$, we find the steady-state equations for $\alpha_{n,0}$ as
\begin{align}
0 =-Dn\alpha_{n,0} + \frac{1}{2}J_{+}S_+(\alpha_{n-1,0} -\alpha_{n+1,0})~,
\label{eq:alphan0}
\end{align}
which are decoupled from all other coefficients $\alpha_{n,m}$ for $m\neq 0$.
With the use of the recurrence relation of the modified Bessel function, it is straightforward to 
find the solution of Eq.~\eqref{eq:alphan0} as
\begin{align}
\alpha_{n,o}=\frac{I_{|n|}\left(\frac{J_+S_+}{D}\right)}{I_0\left(\frac{J_+S_+}{D}\right)}~,
\label{eq:alphan0s}
\end{align}
where $I_n$ is the modified Bessel function. The order parameter value
$S_+$ can be determined self-consistently by the above equation for $n=1$ with
$\alpha_{1,0}=S_+$. Note that $S_+$ depends only on $J_+/D$, independent of $J_-/D$. 
For small $S_+$, we obtain 
\begin{align}
S_+\approx \sqrt{2(J_+/2D-1)},
\end{align}
implying that
the phase wave state is physically meaningful only for $J_+/D \ge 2$.
All other coefficients are set to zero, which satisfy the steady-state condition,
resulting in $\rho^s_{\textrm{(b)}}=\frac{1}{(2\pi)^2} e^{\frac{J_+S_+}{D}\cos(X-\Phi_+)}/I_0(\frac{J_+S_+}{D})$.
There exists also the symmetric solution with $S_+=0$.

\begin{figure}
   \centering
    \includegraphics[width=0.85\columnwidth]{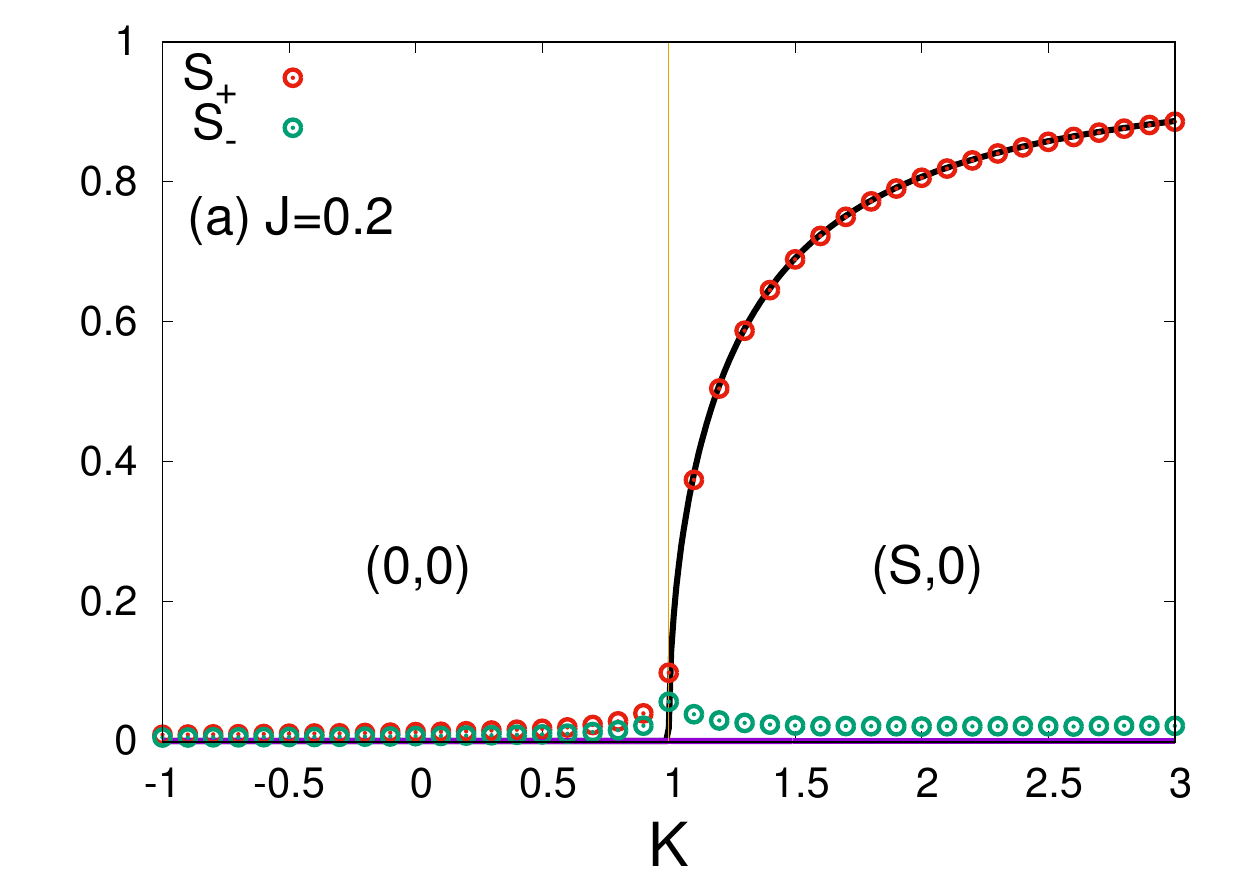}
    \includegraphics[width= 0.85\columnwidth]{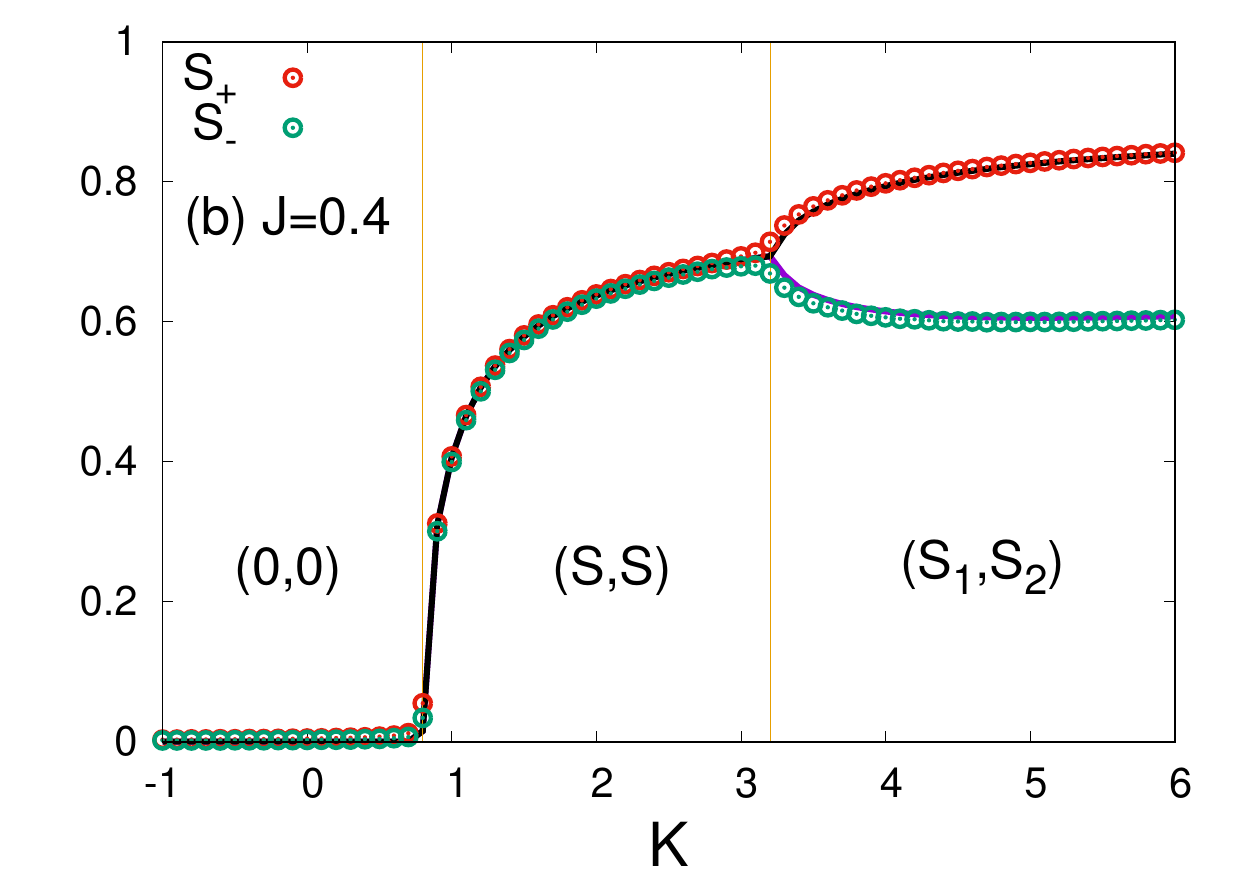}
    \includegraphics[width=0.85\columnwidth]{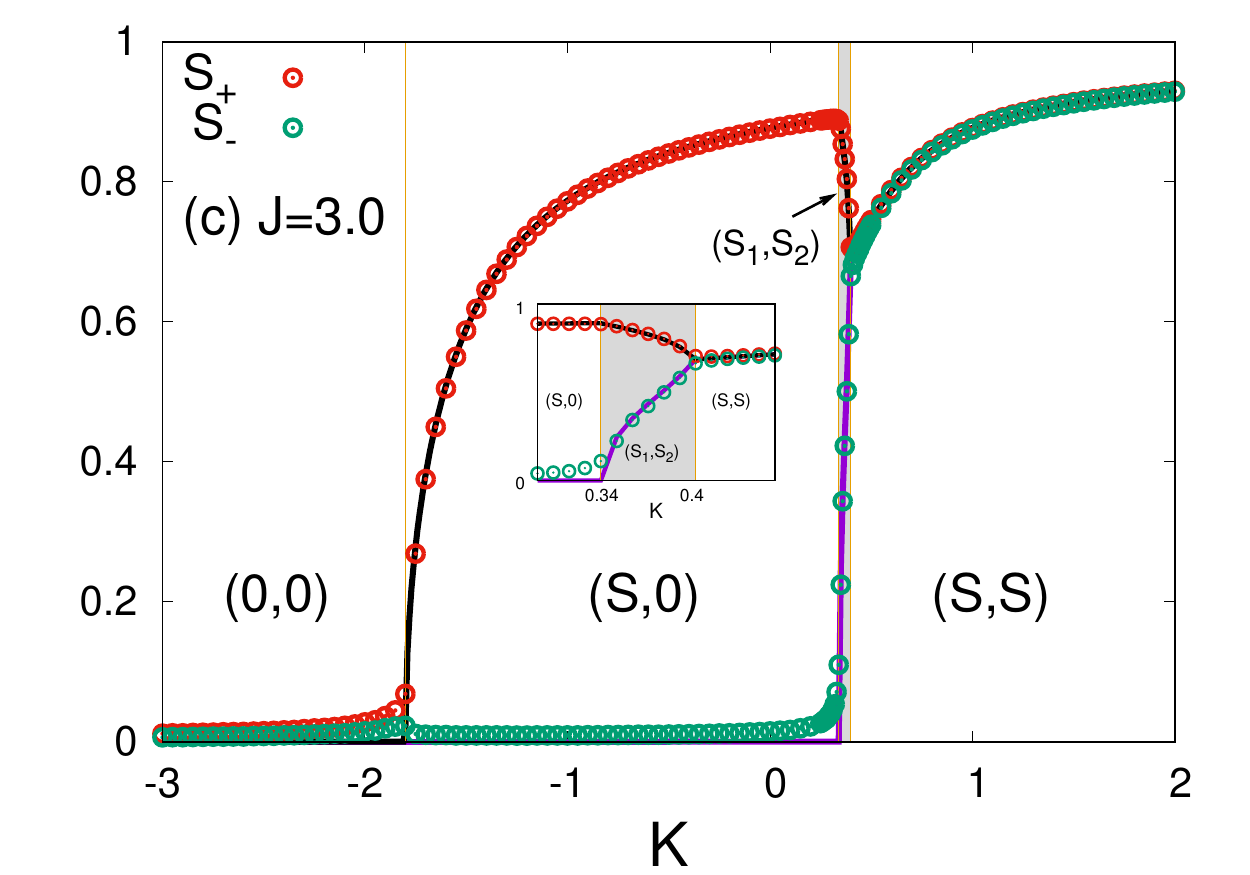}
    \includegraphics[width= 0.85\columnwidth]{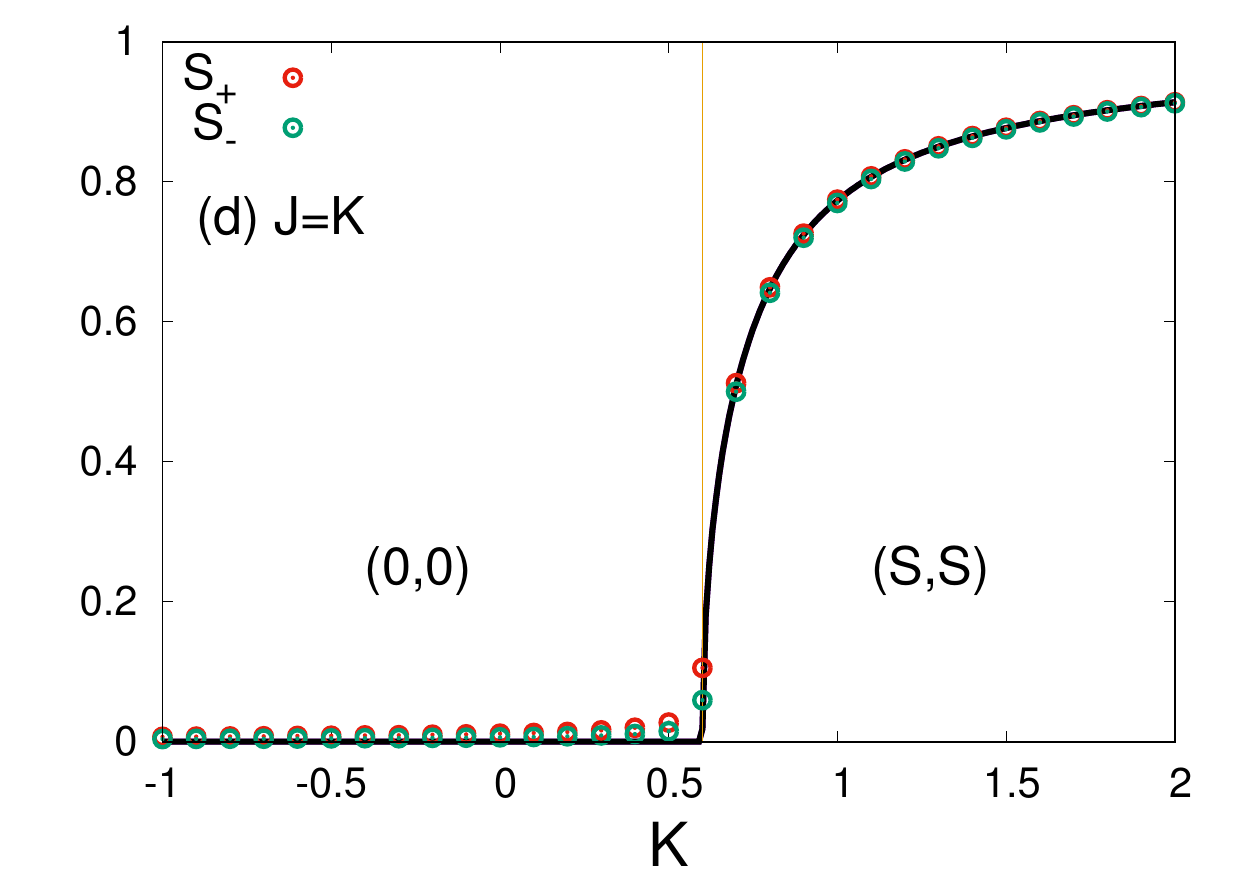}
    \caption{(Color online) Plots of $S_{\pm}$ as a function of $K$ for (a) $J=0.2$; (b) $J=0.4$; (c) $J=3.0$; (d) $J=K$, respectively,
    with temperatures $D_{x}=D_{\theta}=0.15$.
The region for the $(S_1, S_2)$ mixed state in (c) is enlarged and shown in more detail in the inset figure.
Symbols represent the data obtained from
numerical simulations  with $N=10^4$, and the black (purple)
solid line in each panel is given by the exact solutions and iteration results for $S_+$ ($S_-$) given in Sec.~\ref{sec:sss}.
    }
    \label{fig:S_K}
\end{figure}

The steady-state solutions for the mixed state (c) and the ordered state (d) cannot be derived analytically.
Instead, we  truncate higher-order Fourier modes in
the dynamic equation~\eqref{eq:alphanm} by keeping the modes of $(n,m)$ up to order
$\ell(= |n|+|m|)$. Then, steady-state solutions are calculated by numerical iterations
with $\dot\Phi_{\pm}=0$. The solutions well converge with increasing $\ell$ and saturate around $\ell=10$.
This fast convergence is due to the exponential decay of $\alpha_{n,m}$ with $\ell$ in the steady state.
These iteration results are shown by the black solid lines ($S_+$) and the purple ones ($S_-$) in Fig.~\ref{fig:S_K}~(a)-(d)
for various values of $J$ and $K$ with $D_x=D_\theta=0.15$. Both the mixed state and the ordered state are well
established and the exact solution of the phase wave state matches with numerical results perfectly well.
We note that all coefficients $\alpha_{n,m}$ turn out to be real in all steady states, even if one starts from
a complex initial state.

In the EQ situation ($J_-=0$ and $d=0$), the exact solution is possible due to decoupling of the dynamic equations,
yielding $S_+=S_-\equiv S$, which has the same value of $S_+$ in the phase wave state.
The steady-state PDF is given by a simple product of those of the phase wave states for $X$ and for $Y$, and 
the steady-state solutions are given by 
\begin{align}
\alpha_{n,m}=\frac{I_{|n|}\left(\frac{JS}{D}\right)I_{|m|}\left(\frac{JS}{D}\right)}{[I_0 \left(\frac{JS}{D}\right)]^2}~.
\label{eq:alphanms}
\end{align}

Finally, we remark that the vanishing Kuramoto order parameters $R_x$ and $R_\theta$ in Eqs.~\eqref{eq:Rx} and \eqref{eq:Rtheta}
are guaranteed in our Fourier mode analysis, as $R_x$ and $R_\theta$ correspond to half-integer Fourier
modes such as $\alpha_{\frac{1}{2},\pm\frac{1}{2}}$, which do not exist as a Fourier mode with periodicity of $2\pi$.

\section{linear stability analysis of the disordered state}
\label{sec:lsa}

The disordered (incoherent) state is characterized by $\alpha_{n,m}=0$ for any pair of $(n,m)\neq (0,0)$. We take a small perturbation
to the disordered state such that $\alpha_{n,m}=\epsilon c_{n,m}$ for $(n,m)\neq(0,0)$ with a small parameter $\epsilon$.
From the dynamic equation \eqref{eq:alphanm} with $S_{+}=\epsilon c_{1,0}$ and $S_{-}=\epsilon c_{0,1}$, we find,
up to the linear order in $\epsilon$,
\begin{align}
&\dot{c}_{1,0}=-D\left(1-\frac{J_+}{2D}\right)c_{1,0},\nonumber\\
 &\dot{c}_{0,1}=-D\left(1-\frac{J_+}{2D}\right)c_{0,1},\\
 &\dot{c}_{n,m}=-\left[D(n^2+m^2)+2dnm\right] c_{n,m}~ \textrm{for other}~(n,m). \nonumber
\end{align}
Thus, the stability condition is simply given by $J_+/D <2$ as 
$D(n^2+m^2)+2dnm=D_x(n+m)^2+D_\theta(n-m)^2 >0$. As the other steady states are defined only for
$J_+/D>2$, the disordered state should be globally stable for $J_+/D<2$, which agrees perfectly well
with numerical data obtained by iterations of the dynamic equation \eqref{eq:alphanm}.

\section{Numerical simulations}
We perform numerical
simulations by integrating Eqs.~\eqref{eq:X} and \eqref{eq:Y} for various values of $J$ and $K$
with temperatures $D_x=D_\theta=0.15$ for $N=10^4$.
We used the Euler method with step size $\delta t =0.01$ for $M_t=10^5$ time steps, where the initial $M_t/2$ steps of each run were discarded as a transient, after which $S_+$ and $S_-$  were measured and averaged over remaining time for 10 independent samples.



Figure~\ref{fig:S_K} shows simulation data (symbols) for $S_{\pm}$ versus $K$ when $J=0.2$, 0.4, 3.0 and when $J=K$.
All data are in full agreement with exact solutions and iteration results in Sec.~\ref{sec:sss}.
Note that the transition from the disordered state to other steady states should occur
at $J_+=2D$, derived as the instability threshold in the preceding section. With $D=0.3$, this transition point 
is located at $J+K=4D=1.2$, which are fully consistent with data. Small finite-size effects in simulation data can be seen  near
the transition from the disordered state.

\section{Phase diagram}\label{sec:pd}
\begin{figure}
   \centering
    \includegraphics[width=1.0\columnwidth]{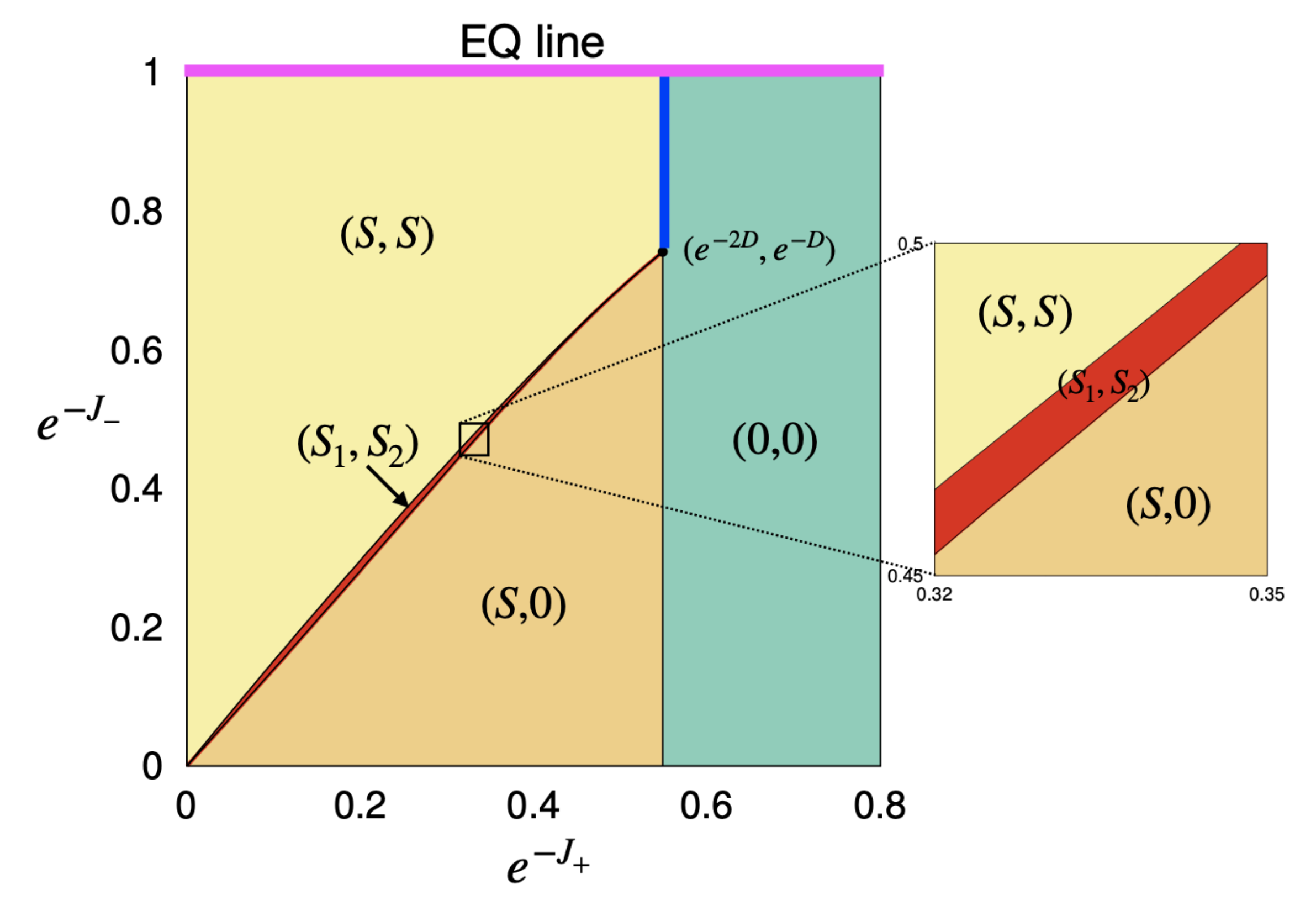}
    \caption{(Color online) Phase diagram in the $(e^{-J_{+}}, e^{-J_{-}})$ plane
    with $D_x=D_\theta=0.15$ ($D=0.3$ and $d=0$). Each phase is indicated by a distinct color.
    The mixed phase colored in red is enlarged in the inset for better visibility.
    The disordered phase becomes unstable at $J_+=2D$ and
    all four phases converge at a single multicritical point with its coordinate of $(e^{-2D}, e^{-D})$.
    %
The pink line ($J_-=0$) represents the equilibrium (EQ) line, corresponding to the two decoupled XY models
with the magnetic transition at $J_+=2D$. It is noteworthy that
direct transitions are possible
from the disordered phase to the ordered phase through the blue solid line in non-equilibrium situations.
}
  \label{fig:phd}
\end{figure}

Figure~\ref{fig:phd} presents the phase diagram in the $(e^{-J_{+}}, e^{-J_{-}})$ plane for convenience,
when the temperatures are equal ($D_x=D_\theta$, i.e.~$d=0$). In this case, the dynamic equations \eqref{eq:x} and \eqref{eq:theta} 
are symmetric under the interchange of $x\leftrightarrow\theta$ and $K\leftrightarrow J$. Thus, the phase diagram 
for $e^{-J_-}>1$ can be easily deduced from that for $e^{-J_-}<1$. In order to locate the phase boundary more precisely,
we use the exact solutions and the iteration results with terms up to the order $\ell=|n|+|m|=20$.  

\begin{figure}
   \centering
    \includegraphics[width=0.9\columnwidth]{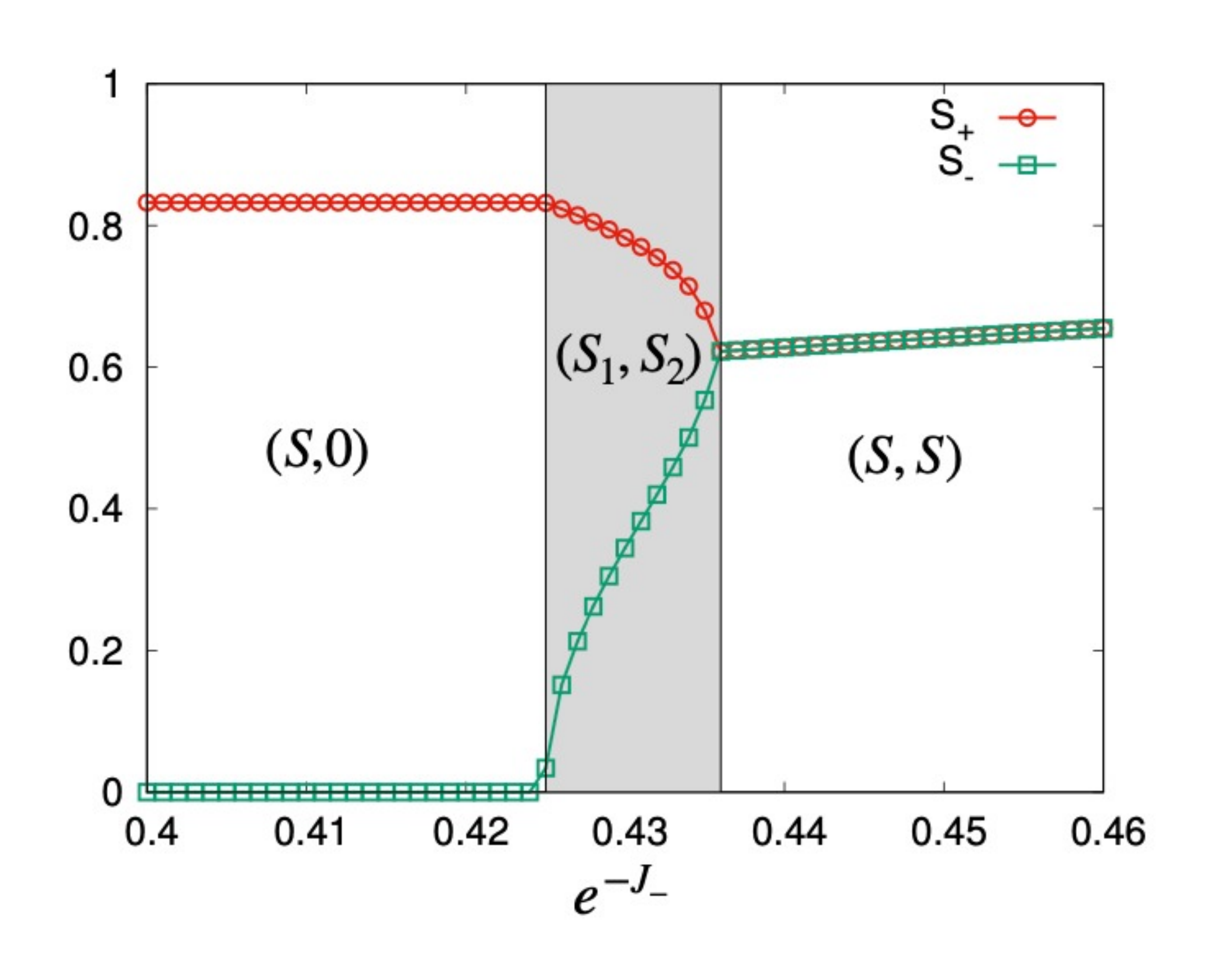}
    \caption{(Color online) Plots of $S_{\pm}$ versus $e^{-J_{-}}$ for $e^{-J_{+}}=0.3$.
    Symbols represent the data obtained from iteration results with $\ell=20$, and the lines are 
guide to the eyes.
}
  \label{fig:S_expmJpJm}
\end{figure}

The phase diagram reveals several noteworthy features.
First, the mixed phase $(S_1,S_2)$ colored in red clearly exists and separates the phase wave phase $(S,0)$ and 
the ordered phase $(S,S)$ completely, though the mixed state region itself is quite small in size. 
In Fig.~\ref{fig:S_expmJpJm}, the order parameter values are plotted along a vertical line in the phase diagram at $e^{-J_+}=0.3$ 
in the range of   $0.40 \leq e^{-J_{-}} \leq 0.46$. The value of $S$ remains constant in the phase wave phase
$(S,0)$, as predicted by the exact solution in Sec.~\ref{sec:sss}. This value is also observed at the EQ point ($e^{-J_-}=1$)
in the ordered phase. As $e^{-J_-}$ is increased, the system goes through the mixed and the ordered phase in succession before 
arriving at the EQ point.

Second, the disordered phase $(0,0)$ appears for $J_+/D < 2$ for all values of $J_-$, consistent with
the linear stability analysis result presented in Sec.~\ref{sec:lsa}.
Third, all four phases converge at a single multicritical point where $J_+/D=2$ and $J_-/D=1$. The value of $J_-$ at the
multicritical point will be derived exactly through a perturbation theory near the disordered state in the following section.

Finally, we point out that direct transitions from the disordered phase to the ordered phase are possible 
not only along the EQ line but also at $J_+=2D$ (blue line in Fig.~\ref{fig:phd}) in nonequilibrium situations.
This feature represents a major difference from the system with quenched disorder only~\cite{ref:1Dwithfrequency},
where the direct transition is possible only at the symmetric point $J=K$ ($e^{-J_-}=1$).
It would be of interest to study the existence and location of a multicritical point when both quenched
and thermal disorder are present simultaneously, which is left for future study.


We also studied the case with two different temperatures ($D_x\neq D_\theta$). 
In this case, the dynamic equations are symmetric with 
$D_x\leftrightarrow D_\theta$, in addition to $x\leftrightarrow\theta$ and $K\leftrightarrow J$.
The qualitative features of the phase diagram are similar to those depicted in Fig.~\ref{fig:phd}. 
We find the multicritical point, the $J_-$ value of which is given by $J_-=D+d$, while 
the $J_+$ value remains unchanged at $2D$ (see Sec.~\ref{sec:pt}).





\section{perturbation theory near the disordered state}
\label{sec:pt}
For convenience, we rewrite the mode-coupled equations \eqref{eq:alphanm}
with $\dot\Phi_{\pm}=0$ as
\begin{align}
&{\dot{\alpha}}_{n,m} =-[u(n^2+m^2)+2znm]\alpha_{n,m}\nonumber\\
&\quad\quad+ (n+vm)\alpha_{1,0}(\alpha_{n-1,m} -\alpha_{n+1,m})
\nonumber\\
&\quad\quad+(vn+m) \alpha_{0,1}(\alpha_{n,m-1}-\alpha_{n,m+1})
\label{eq:alphanm1}
\end{align}
with new parameters 
\begin{align}
u=2D/J_+, \quad v=J_-/J_+, \quad \textrm{and} \quad z=2d/J_+ ~,
\end{align}
where time $t$ is scaled by a factor of $2/J_+$ for simplicity.
We consider $u, v \ge 0$ only, because only the trivial disordered phase is possible for $u<0$
and there is a reflection symmetry under $v\leftrightarrow -v$. Also the positivity of diffusion
constants ($D_x >0, D_\theta>0$) demands $|z|<u$.

As the ordering begins to emerge for $u\lesssim 1$,
we introduce a small positive parameter $\delta$ such that $u=1-\frac{\delta}{2}$.
From the exact relation of Eq.~\eqref{eq:alphan0s} for the phase wave state, we obtain the steady-state values 
for the Fourier coefficients $\alpha_{n,0}$
in lower orders of $\delta$ as
\begin{align}\label{eq:pw}
\alpha_{n,0}\approx \frac{\delta^{|n|/2}}{|n|!}
\left(1+\frac{|n|(|n|-2)}{3(|n|+1)}\delta+O(\delta^2)\right)~,
\end{align}
where we utilized the expansion of $I_n(a)$ for small $a$ 
and $S_+^2\approx \delta (1-\delta/3+\delta^2/72)+O(\delta^3)$.
Along the EQ line, it is also straightforward to obtain from Eq.~\eqref{eq:alphanms}
\begin{align}
\alpha_{n,m}=\alpha_{n,0}\alpha_{0,m}\approx \frac{\delta^{(|n|+|m|)/2}}{|n|!|m|!}\left(1+O(\delta)\right)~,
\end{align}
which demonstrates the rapid exponential decay of the
steady-state Fourier coefficients with $\ell=|n|+|m|$.
These analytic results motivate us to assume that  $\alpha_{n,m}$ in lower orders 
for general steady states can be expressed as
\begin{align}
\alpha_{n,m}\approx {\delta^{(|n|+|m|)/2}}\left(a_{n,m}+b_{n,m}\delta\right)~,
\label{eq:deltanm}
\end{align}
with $a_{0,0}=1$ and $b_{0,0}=0$ for normalization and 
$a_{n,m}=a_{-n,-m}$ and $b_{n,m}=b_{-n,-m}$.

\subsection{steady states}
Insert this leading order into Eq.~\eqref{eq:alphanm1} with ${\dot{\alpha}}_{n,m}=0$ 
for  steady states, we can then determine $a_{n,m}$ order by order in $\delta$.
Up to the order of $\delta^{3/2}$, we obtain relations between $a_{n,m}$ by
\begin{align}
&a_{2,0}=\frac{1}{2} a_{1,0}^2~, \qquad\qquad~~ a_{0,2}=\frac{1}{2} a_{0,1}^2~, \\
&a_{1,1}=\frac{1+v}{1+z}a_{1,0} a_{0,1}~, \quad a_{1,-1}=\frac{1-v}{1-z}a_{1,0} a_{0,1}~,\\
&a_{1,0}-2a_{1,0}a_{2,0}+2v a_{0,1}[a_{1,-1}-a_{1,1}]=0~,\\
&a_{0,1}-2a_{0,1}a_{0,2}+2v a_{1,0}[a_{1,-1}-a_{1,1}]=0~,\\
&a_{3,0}=\frac{1}{3}a_{1,0}a_{2,0}, \qquad\qquad a_{0,3}=\frac{1}{3}a_{0,1}a_{0,2}~,\label{eq:30}\\
&a_{2,1}=\frac{1}{5+4z}\left[ (2+v) a_{1,0}a_{1,1}+(1+2v)a_{0,1}a_{2,0}\right],\\
&a_{1,2}=\frac{1}{5+4z}\left[ (2+v) a_{0,1}a_{1,1}+(1+2v)a_{1,0}a_{0,2}\right],\\
&a_{2,-1}=\frac{1}{5-4z}\left[ (2-v) a_{1,0}a_{1,-1}+(1-2v)a_{0,1}a_{2,0}\right],\\
&a_{1,-2}=\frac{1}{5-4z}\left[ (2-v) a_{0,1}a_{1,-1}+(1-2v)a_{1,0}a_{0,2}\right]~.
\end{align}
There are also relations for $b_{1,0}$ and $b_{0,1}$, which are shown in the Appendix~\ref{appendix:relations}.

By combining the first four equations, four distinct solutions can be obtained, each of which represents 
a fixed point such as $(a_{1,0}, a_{0,1})=$ $(0,0)$, $(1,0)$, $(0,1)$, $(\sqrt{a},\sqrt{a})$
with $a= [1+4v(v-z)/(1-z^2)]^{-1}$. The $(0,0)$ fixed point is associated with the disordered phase (a), while
the $(1,0)$ and $(0,1)$ points denote two equally probable states of the phase wave phase (b).
The $(\sqrt{a},\sqrt{a})$ point corresponds to the symmetric ordered phase (d). 
All other coefficients $a_{n,m}$ up to $\ell=3$ can be calculated by inserting the fixed point solutions into the above equations.
Up to this order, the mixed phase (c) does not appear.

We also report the fixed point values of $(b_{1,0}, b_{0,1})$, which represent the $O(\delta^{3/2})$ term of 
$(\alpha_{1,0}, \alpha_{0,1})$ as $(0,0)$, $(-1/6,0)$, $(0, -1/6)$, and $(b,b)$ for four fixed points, respectively.
The value of $b_{n,m}$ is discussed in the Appendix~\ref{appendix:relations}.
 All other terms are higher-order than $O(\delta^{3/2})$, thus can be set to zero.

\subsection{linear stability analysis of the phase wave state}

The steady state solution of the phase wave state is quite simple as
$a_{1,0}=1, a_{2,0}=1/2, a_{3,0}=1/6, b_{1,0}=-1/6$ with all other
coefficients $a_{n,m}=0$, up to the order of $\delta^{3/2}$. In order to check its stability, we take a small perturbation to this state 
by adding $\epsilon c_{n,m}$ to the steady state values of $\alpha_{n,m}$ 
except for $(0,0)$. Then, it is straightforward to find up to the linear order in $\epsilon$  
from Eq.~\eqref{eq:alphanm1} as
\begin{align}
&\dot{c}_{1,0}=- c_{2,0}\sqrt{\delta}~,\label{eq:c10}\\
&\dot{c}_{0,1}=\frac{1}{2}c_{0,1}\delta +v(c_{1,-1}-c_{1,1})\sqrt{\delta}~,\label{eq:c01}\\
&\dot{c}_{2,0}=-4c_{2,0}+2[2c_{1,0}-c_{3,0}]\sqrt{\delta}~,\\
&\dot{c}_{0,2}=-4 c_{0,2}+2v[c_{1,-2}-c_{1,2}]\sqrt{\delta}~,\\
&\dot{c}_{1,1}=-2(1+z)c_{1,1}+(1+v)[2c_{0,1}-c_{2,1}]\sqrt{\delta}~,\label{eq:c11}\\
&\dot{c}_{1,-1}=-2(1-z)c_{1,-1} +(1-v) [2c_{0,1}-c_{2,-1}]\sqrt{\delta},\label{eq:c1m1}\\
&\dot{c}_{n,m}=-(n^2+m^2+2znm)c_{n,m} \\
&\quad+ (n+vm)(c_{n-1,m}-c_{n+1,m})\sqrt{\delta} ~~ \textrm{for other}~(n,m)\nonumber
\end{align}
in the lower order of $\delta$.

Note that most of $c_{n,m}$ vanish in the long time limit for small $\delta$, as
$n^2+m^2+2znm >0$. In contrast, $c_{0,1}$ may diverge depending on details of Eq.~\eqref{eq:c01},
thus $c_{1,1}$ and $c_{1,-1}$  may also diverge as their dynamics involve $c_{0,1}$, as 
in Eqs.~\eqref{eq:c11} and \eqref{eq:c1m1}. Consequently, we can drop vanishing terms as $c_{3,0}$, $c_{1,-2}$, $c_{1,2}$,
$c_{2,1}$, and $c_{2,-1}$ in Eqs.~\eqref{eq:c10}-\eqref{eq:c1m1}, leading to the closed equations for
$c_{1,0}$, $c_{0,1}$, $c_{2,0}$, $c_{0,2}$, $c_{1,1}$, and $c_{1,-1}$. These linear equations can be analyzed by 
the standard eigenvalue analysis, yielding five negative eigenvalues and one eigenvalue
$\lambda=[(2v-z)^2-1]\delta$. Hence, the stability condition is given by
\begin{align}
|2v-z|<1~,
\end{align}
implying $|J_- - d|<\frac{1}{2}|J_+|$ in terms of original parameters, which 
predicts the location of the multicritical point precisely as $J_-=D+d$ with $J_+=2D$ 
(see Fig.~\ref{fig:phd} and Sec.~\ref{sec:pd}).

\section{Examples of real word swarmalators}
In this section, we list some examples of real world swarmalators which exhibit some of the collective states we have found (see \cite{ref:review1} for an exhaustive list of real world swarmalators). 
The async(disordered state), sync(ordered state), and phase wave states have 
all been observed in these systems. The mixed state has not been \textit{explicitly} observed, but due to its close resemblance to the phase wave, it may in fact have been realized but not observed. 

(i) {\it{Vinegar eels}} are a type of nematode (a family of worm microswimmers) commonly found in beer mats and tree slime \cite{ref:vinegareels}. When suspensions of these eels are prepared on glass disks they swarm around the 1D edges and synchronize the beating of their tails, thereby satisfying the defintion of a 1D swaramlators. Under certain conditions they form metachronal waves, which are equivalent to the phase waves states here found but winding number
$k > 1$~\cite{ref:vinegareels}.

(ii) {\it{Sperm}} is another type of microswimmer which can be considered a 1D swarmalator. When sperm from ram semen is confined to 1D circular geometeries,
they show a transition from an isotropic state (equivalent to our async state)
to an uniformly rotating vortex (equivalent to our phase wave state)~\cite{ref:sperms}.
Sperms are free to move in 2D and also form synchronous clusters.

(iii) {\it{C. elegans}} are another type of microswimmers which also swarm and
sync the gait of their tails.  When confined to 1D channels they form
synchronous clusters, which is analogous to the sync state~\cite{ref:yuan2014}.

(iv) {\it{Magnetic domain walls}} are characterized by a center of mass $x$ and a magnetic dipole vector with orientation $\theta$. When subject to external forcing, both $x$ and $\theta$ undergo periodic motion as so the walls can be considered swarmalators~\cite{ref:magneticdomainwalls}. In an experiment with $N = 2$ walls, different varieties of sync phenomena can be observed.

(v) {\it{Bristle bots}} are homemade `automata' made from a toothbrushes and powered by a simple cell-phone motor \cite{ref:bristle}.
When placed on a circular drum, they self-organize around the ring-like edge.
The incoherent state is in ~Fig.~9~a, phase wave state (Fig.~9~b), and
the sync state (Fig.~9~c) are observed (By Fig.~9 we mean Fig.~9 in
Ref.~\cite{ref:bristle}).

\bibliographystyle{ieeetr}


\section{Discussions}
The joint action of swarming and synchronization defines a new type of emergence
about which little is known.  The paper is part of effort to explore this unchartered terrain
by studying a simple 1D model of swarmalators subject to thermal noise.
We found four collective states, encapsulated their stabilties in a phase diagram,
illustrated their transitions with order parameter curves $S_{\pm}(J,K)$, and
discussed realizations in nature.
These states were previously reported in a study of the 1D swarmlator model with
quenched disorder~\cite{ref:1Dwithfrequency}, but the bifurcation structure
for the thermal noise is different:
As shown in the phase diagram in Sec.~IX,
through the blue solid line, the phase transition directly occurs
from the $(0,0)$ state to the $(S,S)$ state without going through the $(S,0)$ or
$(S_1,S_2)$, which is very different from the 1D model with quenched
disorder~\cite{ref:1Dwithfrequency}.
In other words, in the absense of thermal noise, there is
only one ``tetracritical'' point there, which means that the
direct phase transition from the $(0,0)$ to the $(S,S)$ state occurs only at
that point.
However, when the thermal noise comes in the system, such region for the direct phase
transition (same as that for the EQ line) is found to enlarge, not only at the multicrossing point,
which is induced by the thermal noise.

We here considered identical swarwalators with same natural frequency, and showed
that the four states can be induced by means of only the thermal noise and
nonequilibrium interactions.
In other words, the nonequilibrium features such as the four states stem from the
thermal noise and the interaction between the units in the system.
We here claim that the quenched disorder is not the inevitable component to induce the four states.
Specifically, we have unequivocally detected the presence of the mixed state in the system.
We have conducted a stability analysis of the phase wave state using the perturbation theory, and determined the 
multicritical point where the four states converge. 
Our findings demonstrate a great level of agreement with numerical results, indicating the originality of 
our work.  

The main theoretical takeaway of our work that allowing oscillators to swarm
greatly enriches their collective behavior.  The Kuramoto model of regular oscillators
with thermal noise, for example, has a single transition: from incoherence to
synchrony~\cite{ref:footnote}.
The 1D swarmalator model on the other hand has four collective states and four
distinct transitions, as depicted on Fig.~\ref{fig:S_K}.  An interesting implication of
this novel bifurcation structure is that synchrony does not increase monontonically
with the phase coupling $K$ (as happens in the Kuramoto model). Holding $J$ constant
and turning up $K$ can either leave you stuck in the phase wave (Fig~\ref{fig:S_K}(a))
or take you \textit{out} of the sync state into the
mixed state (Fig~\ref{fig:S_K}(b)) -- in both cases the overall phase synchrony decays.

An tantalizing goal for future work would be to find the mixed state in a realworld
system of swarmalators. While to our knowledge it has not directly reported, we suspect
the mixed state might be lurking within the recently reported metachronal waves of
vinegar eels~\cite{ref:vinegareels}. As we mentioned, metachronal waves are mimicked by our phase wave;
since the phase wave is visually similar to the mixed state, the phase wave might have been
misidentified. If experimenters could extract the $(x_i, \theta_i)$ of each eels,
then the $S_{\pm}$ could be measured and in principle the mixed state could be detected.
The search for the mixed state could also include other biological microwswimmers,
magentic domain walls, or bristlebots. There are also theoretical avenues to
explore in the future. One could study the effects of delayed coupling,
external forcing, or mixed sign interactions.

\section{Acknowledgements}
This research was supported by NRF Grants No.~2021R1A2B5B01001951 (H.H) and No.~2017R1D1A1B06035497 (H.P.), 
and individual KIAS Grants No.~PG064901 (J.S.L.), and No.~QP013601 (H.P.) at Korea Institute for Advanced Study.

\appendix
\section{relations for higher-order coefficients}
\label{appendix:relations}
Relations for higher-order coefficients are shown as
\begin{align}
&b_{2,0}=a_{1,0}b_{1,0}+\frac{1}{2} \left[a_{2,0}-a_{1,0}a_{3,0}+v a_{0,1} (a_{2,-1}-a_{2,1})\right]~,\nonumber\\
&b_{0,2}=a_{0,1}b_{0,1}+\frac{1}{2} \left[a_{0,2}-a_{0,1}a_{0,3}+v a_{1,0} (a_{1,-2}-a_{1,2})\right]~,\nonumber\\
&b_{1,1}=\frac{1}{1+z}\left[(1+v)(a_{1,0}b_{0,1}+a_{0,1}b_{1,0})\right.\nonumber\\
&\left.\qquad\qquad+\frac{1}{2}\{a_{1,1}-(1+v)(a_{1,0}a_{2,1}+a_{0,1}a_{1,2})\}\right]~,\nonumber\\
&b_{1,-1}=\frac{1}{1-z}\left[(1-v)(a_{1,0}b_{0,1}+a_{0,1}b_{1,0})\right.\nonumber\\
&\left.\qquad\qquad+\frac{1}{2}\{a_{1,-1}-(1-v)(a_{1,0}a_{2,-1}+a_{0,1}a_{1,-2})\}\right]~,\nonumber\\
&b_{1,0}-2(a_{1,0}b_{2,0}+a_{2,0}b_{1,0})\nonumber\\
&\qquad\qquad+2v\left[a_{0,1}(b_{1,-1}-b_{1,1})+b_{0,1}(a_{1,-1}-a_{1,1})\right]=0~,\nonumber\\
&b_{0,1}-2(a_{0,1}b_{0,2}+a_{0,2}b_{0,1})\nonumber\\
&\qquad\qquad+2v\left[a_{1,0}(b_{1,-1}-b_{1,1})+b_{1,0}(a_{1,-1}-a_{1,1})\right]=0~.\nonumber
\end{align}

For the fixed point $(a_{1,0}, a_{0,1})=(1,0)$, we obtain $a_{3,0}=1/6$ and $a_{0,3}=0$ 
from Eq.~\eqref{eq:30}
and all other coefficients are zero. With these values, it is easy to derive
that $b_{1,0}=-1/6$ and all other $b_{n,m}=0$. Note that this value is consistent with Eq.~\eqref{eq:pw}.

In  the ordered phase with $(a_{1,0}, a_{0,1})=(\sqrt{a},\sqrt{a})$, the above equations
for $b_{n,m}$ possess a symmetry under the exchange of indices ($n\leftrightarrow m$) and also linearity,
implying that $b_{1,0}=b_{0,1}$, $b_{2,0}=b_{0,2}$, and $b_{1,1} (v,z)=b_{1,-1} (-v,-z)$. 
This observation excludes the possibility of the mixed phase ($S_+\neq S_-$), at least up to this order. 
It is straightforward to get explicit expressions for $b_{n,m}$ for the symmetric ordered phase, 
but they are too complicated to be displayed here.


\begin{thebibliography}{99}
\bibitem{ref:swarmalators}
K. P. O’Keeffe, H. Hong, S. H. Strogatz, Nature comm. {\bf{8}}, 1 (2017).

\bibitem{ref:sperms}
A. Creppy, F. Plourabou{\'e}, O. Praud, X. Druart, S. Cazin, H. Yu, and P. Degond, J. of The Royal Soc. Interface, {\bf{13}}, 20160575 (2016).

\bibitem{ref:vinegareels}
A. Peshkov, S. McGaffigan, and A. C. Quillen, Soft Matter, {\bf{18}}, 1174 (2022); 
A. C. Quillen, A. Peshkov, E. Wright, and S. McGaffigan, Phys. Rev. E {\bf{104}}, 014412 (2021).

\bibitem{ref:livingcrystals}
T. H. Tan, A. Mietke, J. Li, Y. Chen, H. Higinbotham, P. J. Foster,
S. Gokhale, J. Dunkel, and N. Fakhri, Nature {\bf{607}} 287 (2022).

\bibitem{ref:Japanesefrogs} I. Aihara, T. Mizumoto, T. Otsuka, H. Awano, K. Nagira,
H. G. Okuno, and K. Aihara, Sci. Rep. {\bf{4}}, 1 (2014); K. Ota, I. Aihara,
and T. Aoyagi, Royal Society open science {\bf{7}}, 191693 (2020).

\bibitem{ref:Janusparticles}
J. Yan, M. Bloom, S. C. Bae, E. Luijten, and S. Granick, Nature {\bf{491}}, 578 (2012).

\bibitem{ref:Quinckerollers}
B. Zhang, A. Sokolov, and A. Snezhko, Nature comm. {\bf{11}}, 4401 (2020).

\bibitem{ref:robots}
A. Barci{\'s} and C. Bettstetter, IEEE {\bf{8}}, 218752 (2020).

\bibitem{ref:chemo}
D. Tanaka, Phys. Rev. Lett. {\bf{99}}, 134103 (2007).

\bibitem{ref:diverse}
M. Iwasa and D. Tanaka, Phys. Lett. A {\bf{381}}, 3054 (2017).

\bibitem{ref:pinning}
G. K. Sar, D. Ghosh, and K. P. O’Keeffe, arXiv:2211.02353 (2022).

\bibitem{ref:forcing}
Joao U. F. Lizarraga1, and Marcus A. M. de Aguiar, Chaos {\bf{30}}, 053112 (2020).

\bibitem{ref:delay}
N. Blum, A. Li, K. P. O’Keeffe, and O. Kogan, arXiv:2210.11417 (2022).

\bibitem{ref:finiterange}
H. K. Lee, K. Yeo, and H. Hong, Chaos {\bf{31}} 033134 (2021).

\bibitem{ref:stochastic}
U. Schilcher, J. F. Schmidt, A. Vogell, and C. Bettstetter, IEEE International Conference on Autonomic Computing and Self-Organizing Systems (ACSOS). IEEE, 2021.

\bibitem{ref:coupling}
H. Hong, K. Yeo, and H. K. Lee. Phys. Rev. E {\bf{104}}, 044214 (2021).

\bibitem{ref:gourab}
G. K. Sar, S. N. Chowdhury, M. Perc, and D. Ghosh, New J. of Phys. {\bf{24}}, 043004 (2022).

\bibitem{ref:adv}
T. A. McLennan-Smith, D. O. Roberts, H. S. Sidhu,
Phys. Rev. E {\bf{102}} 032607 (2020).

\bibitem{ref:pre-coupling}
K. P. O'Keeffe and H. Hong Phys. Rev. E {\bf{105}}, 064208 (2022).

\bibitem{ref:ring}
K. P. O'Keeffe, J. H. M. Evers, and T. Kolokolnikov, Phys. Rev. E {\bf{98}}, 022203 (2018).

\bibitem{ref:jimenz}
F. J. -Morales, Phys. Rev. E {\bf{101}}, 062202 (2020).

\bibitem{ref:mfl}
S.-Y. Ha {\textit{et. al.,}}
Kinetic and Related Models {\bf{14}}, 429 (2021).

\bibitem{ref:topological}
P. Degond, D. Antoine, and W. Adam,
arXiv:2205.15739 (2022).

\bibitem{ref:ha}
S.-Y. Ha, {\textit{et. al.,}}
Mathematical Models and Methods in Applied Sciences {\bf{29}}, 2225 (2019).

\bibitem{ref:steven}
S. Ceron, K. P. O'Keeffe, and K. Petersen,
arXiv:2211.06439 (2022).

\bibitem{ref:review1}
K. P. O'Keeffe and C. Bettstetter,
Proc. SPIE 10982, Micro- and Nanotechnology Sensors, Systems, and Applications XI, 109822E (13 May 2019); https://doi.org/10.1117/12.2518682

\bibitem{ref:review2}
Sar, Gourab Kumar Kumar, and Dibakar Ghosh,
Europhys. Lett. (2022).

\bibitem{ref:Ising}
Cipra, Barry A, "An introduction to the Ising model." The American Mathematical Monthly {\bf{94}} 937 (1987).

\bibitem{ref:Kuramoto}
Y. Kuramoto, {\textit{Chemical oscillations, Waves, and Turbulence}} (Springer, Berlin, 1984).

\bibitem{ref:1D}
Kevin O'Keeffe, Steven Ceron, and Kirstin Petersen
Phys. Rev. E {\bf{105}}, 014211 (2022).

\bibitem{ref:1Dwithfrequency}
S. Yoon, K. P. O'Keeffe, J. F. F. Mendes, and A. V. Goltsev, Phys. Rev. Lett. {\bf{129}}, 208002 (2022).

\bibitem{ref:verberck}
B. Verberck, Nature Physics {\bf{18}}, 131 (2022).

\bibitem{ref:XY} J. M. Kosterlitz and D. J. Thouless, J. Phys. C {\bf 6}, 1181 (1972).

\bibitem{ref:OA}
E. Ott and T. M. Antonsen, Chaos, {\bf{18}}, 037113 (2008); {\it{ibid.}}, {\bf{19}}, 023117 (2009).

\bibitem{ref:KS}
H. Sakaguchi, S. Shinomoto, and Y. Kuramoto, Prog. Theor. Phys. {\bf{79}}, 600 (1988); S. -W. Son and H. Hong, Phys. Rev. E {\bf{81}}, 061125 (2010).
\bibitem{ref:HJHP} H. Hong, J. Jo, C. Hyeon, and H. Park, J. Stat. Mech, 074001 (2020).

\bibitem{ref:Pikovsky}
B. Sonnenschein and L. Schimansky-Geier, Phys. Rev. E {\bf{88}}, 052111 (2013);
I. V. Tyulkina, D. S. Goldobin, L. S. Klimenko, and A. Pikovsky,
Phys. Rev. Lett. {\bf{120}}, 264101 (2018).

\bibitem{ref:Vicsek}
D. Levis, I. Pagonabarraga, and B. Liebchen, Phys. Rev. Research {\bf{1}}, 023026 (2019);
B. Ventejou, H. Chat\'e, R. Montagne, and X. -q. Shi, Phys. Rev. Lett. {\bf{127}}, 238001 (2021).

\bibitem{ref:Risken}
H. Risken, The Fokker-Planck Equation, Methods of Solution, and Applications (Springer, Berlin, 1996).







\bibitem{ref:yuan2014}
J. Yuan, D. Raizen, and H. H. Bau, Proc. of the Nat. Acad. of Sci.,
  {\bf{111}}, 6865 (2014).

\bibitem{ref:magneticdomainwalls}
A. Hrabec, V. K{\v{r}}i{\v{z}}{\'a}kov{\'a}, S. Pizzini, J. Sampaio,
A. Thiaville, S. Rohart, and J. Vogel, Phys. Rev. Lett. {\bf{120}}, 227204 (2018); 
E. Haltz, S. Krishnia, L. Berges, A. Mougin, and J. Sampaio, Phys. Rev. B {\bf{103}}, 014444 (2021).



\bibitem{ref:bristle}
L. Giomi, N. Hawley-Weld, and L. Mahadevan, Proc. of the Royal Soc.
https://doi.org/10.1098/rspa.2012.0637

\bibitem{ref:footnote}
This scenario is in fact illustrated Fig~\ref{fig:S_K}(d), because in the special case $J=K$ our model reduced to two independent Kuramoto models.

\end{thebibliography}






\end{document}